\documentclass{llncs}

\usepackage{amsthm,amsmath, amssymb, hyperref}

\usepackage[latin1]{inputenc}
\usepackage{graphicx}
\usepackage[usenames, dvipsnames]{color}
\usepackage{tikz}
\usepackage{mathtools}
\usepackage[colorinlistoftodos,prependcaption]{todonotes}
\usepackage{xfrac}

\theoremstyle{definition}

\theoremstyle{definition}
\newtheorem{rul}{Rule}[]

\begin{document}

\newcommand{\pluseq}{\mathrel{+}=}
\newcommand{\mineq}{\mathrel{-}=}

\pagenumbering{arabic}
\pagestyle{plain}

\title{Generating Synthetic Data for Real World Detection of DoS Attacks in the IoT}
\author{Luca Arnaboldi \and Charles Morisset}
\institute{School of Computing, Newcastle University, Newcastle upon Tyne, UK\\ \{l.arnaboldi,charles.morisset\}@ncl.ac.uk}
\maketitle

\begin{abstract}
Denial of service attacks are especially pertinent to the internet of things as devices have less computing power, memory and security mechanisms to defend against them. 
The task of mitigating these attacks must therefore be redirected from the device onto a network monitor.
Network intrusion detection systems can be used as an effective and efficient technique in internet of things systems to offload computation from the devices and detect denial of service attacks before they can cause harm.
However the solution of implementing a network intrusion detection system for internet of things networks is not without challenges due to the variability of these systems and specifically the difficulty in collecting data. 
We propose a model-hybrid approach to model the scale of the internet of things system and effectively train network intrusion detection systems. Through bespoke datasets generated by the model, the IDS is able to predict a wide spectrum of real-world attacks, and as demonstrated by an experiment construct more predictive datasets at a fraction of the time of other more standard techniques.

\end{abstract}

\section{Introduction}

A Denial of Service (DoS) attack targets the availability of a device or network~\cite{Mirkovic:2004:IDS:1044905}, with the intent of disrupting system usability. 
The most common method is referred to as Flooding DoS~\cite{Mirkovic:2004:IDS:1044905}, and may be used as an attempt to deplete the devices' resources including memory, bandwidth and/or battery.
A DoS attack against an Internet of Things (IoT) network has the potential to be significantly more detrimental than one against a standard network, this increased vulnerability is due in part to the low computational power and battery power characteristic of IoT devices~\cite{suo2012security}.

The extant literature has delineated several potential approaches that may be effective in the mitigation of a DoS attack~\cite{talpade2003mitigating}. They widely speaking fall into two categories, host based (e.g. Client Puzzles) which puts the computational effort on the device and network based (e.g. firewall) which offloads the computational effort to a remote server or more powerful device within the system.
However many of these approaches may not scale well in the IoT as computational power, heterogeneity and the large scale of these systems are all limiting factors that deplete the available choices. 
One approach that sidesteps many of these standard detriments is a  Intrusion Detection System (IDS) bespoke to the IoT system to protect. An IDS is a monitor placed on the network that analyses incoming messages to detect attacks and/or unwanted traffic. They are trained using system behaviour data and use these patterns to make the detection.

Organizations and researchers alike have widely recognised the advantages of adapting IDSs as the norm to monitor against DoS attacks on their systems~\cite{mukkamala2002intrusion}. 
Standard approaches used to train IDSs include using a database of known attacks ({\em misuse detection}) and testing systems to create a ``benchmark" behaviour and flag any anomaly as a potential attack ({\em anomaly detection})~\cite{mell2003overview}.
Implementing an IDS within an IoT network however faces multiple challenges: 
Firstly, it is usually challenging to establish a benchmark behaviour in dynamic IoT systems as devices may constantly shift, new devices might join and behaviours might change~\cite{gubbi2013internet}, which might prevent using anomaly detection; 
Secondly, protocols can vary from one network to another, which necessitates data collection to be bespoke to an individual system~\cite{guillen2015inefficiency}; And  
thirdly, a misuse detection can be time consuming to enforce, since collecting data unique to a system and for each attack is time consuming~\cite{mell2003overview} and some system changes can require data (or part of the data) to be collected from scratch (e.g. interactive smart homes where devices can change frequently). 

To address the second and third challenges, we present a novel modelling approach. In brief, our model is a  Markov Decision Process (MDP), representing the IoT network, the attackers, and some processes monitoring the security metrics under consideration. 
A trace of the model (corresponding to a sequence of actions of the MDP) should match a trace of the actual system, and vice versa, such that it becomes possible to train a IDS for the actual system on the traces of the model. 
The main strengths of our approach is the ability to easily represent various configurations for the IoT network as well as multiple types of attackers.
MDPs have some key advantages: they have substantial tool support such as PRISM Model Checker~\cite{kwiatkowska2002prism}, they rely on probabilities and non-determinism to recreate systems and they provide the ability to find the optimum paths through the system using the reward function. 
Through the reward function we create traces of behaviour that mimic attacks on systems by assigning rewards to successful (damaging) behaviour.
Our results highlighted that through this methodology we were able to consistently produce datasets that resulted in accurate IDSs (detecting attacks on real world systems) and that could be trained in a fraction of the time.
The core contributions of this paper are 1) A model of an IoT system that enables the generation of synthetic data sets of network behaviour 2) Modelling of attack behaviour against a system to train a real world IDS 3) A quantitative analysis and validation of this model against a real world implementation of the same system to validate our methodology. 

The paper is split into the following sections; 
In section~\ref{related-work} we discuss the related work; In section ~\ref{problem} the problem overview is discussed; In section \ref{IoT-system} we introduce our IoT system model and attacks model that generates the network behaviour; In section~\ref{experiments} we highlight our assessment methodology; In section~\ref{setup} we discuss the setup for the experiment; Section~\ref{results} provides an analysis of our results and section~\ref{conclusion} concludes and discusses future work.

\section{Related Work}\label{related-work}

\subsection{DoS Attacks on IoT Systems}

DoS attacks have long been one of the most common and dangerous threats in any internet system. These attacks become even more dangerous as the IoT spreads across a vast amount of spectra and parts of life including safety critical and potentially life endangering ones such as IoT Healthcare and Intelligent Transportation Systems. 

The extant literature highlights several new DoS attacks against IoT system taking advantage of unique qualities and IoT infrastructures~\cite{liang2016denial,roman2013features,buennemeyer2007battery}.
One such attack, battery drain attack focuses on exhausting the devices battery power as replacing it might be costly, difficult and lead to extensive periods of downtime. 
These kinds of attack are very subtle as the behaviour of the attacker might not necessarily mimic more common attacks such as pure flooding, they attempt to find battery intensive operations (not necessarily malicious) and repeat them until the device is out of power. 
This is only one specific example of the literature cited above, however, what all of the above have in common is that they are specialised in their intent of disrupting IoT devices and many of the current detection systems do not account for them~\cite{roman2013features}. 
The literature highlights that there is a constant evolution of attacks, as can be seen using resources such as ExploitDB~\cite{exploitDB}. 
When filtering for IoT attacks we can observe that there is a huge increase in the spectrum of attacks targeting these systems.

These upwards trends in combination with the expansion of the IoT across various field makes a good argument for a simple way to observe the impact of these attacks. 
A formalised model would allow for intuitive means to observe and quantify these attacks as well as better defend these systems by generating network behaviour bespoke to them. 

\subsection{Intrusion Detection Systems}

The growing use of internet services in the past few years have facilitated an increase in DoS attacks. 
Despite the best preventative measures, DoS attacks have been successfully carried out against various companies and organizations enforcing the need for better prevention/detection mechanisms.
This is partially due to the vast new avenues of attack (often unique to IoT) that signature based schemes such as SNORT~\cite{roesch1999snort} struggle to detect. 
Further work attempts a more scalable approach that models behaviour of a network (stationary or non-stationary) and labels abnormal packets as a potential anomaly~\cite{bhuyan2014network}. 
Limitations of this approach are a large number of false positives as well as lack of information regarding the attack (e.g. the specific vulnerability the attack relies on) as opposed to a signature based IDS that is able to tell you exactly what rule is broken.

The approach suggested in this paper allows for a mixture of these approaches tackling the limitations of both works.
By modelling behaviour of a system, one can detect any anomaly similar to the second approach and by modelling various attacks it can also provide accurate data of the system behaviour whilst being targeted, allowing for less false positives. 
To predict ``unknown" attacks, the modelling approach uses a stochastic attacker that attempts different behaviours allowed by the system policy. Using this data it can create a wide range of attack signatures and simulate an attacker probing the system.

\subsection{Modelling IoT Systems}

Several papers address modelling IoT, adopting various different approaches. 
Fruth~\cite{fruth2011formal} examines various properties of a wireless network protocol namely connectivity and energy power through PRISM, including quantifying the battery drainage of certain randomized protocols. In previous work~\cite{arnaboldi17dos} we model basic flooding DoS attacks through PRISM and look at the effectiveness of different attack strains and mitigation techniques in defending systems of interconnected IoT devices. 

Our proposed method combines these approaches to recreate an accurate representation of system behaviour and represent a wide range of DoS attacks. PRISM has been widely used as a excellent method to evaluate and verify models of IoT systems and protocols, combining these two models by adapting both the system models and the attack models we successfully model the behaviour and general properties of a bespoke IoT system. 
We then use the inbuilt verification capabilities to ensure correctness relative to mimicking system behaviour by establishing benchmarks and tests. PRISM and its inbuilt simulation capabilities allows to simulate attacks against the verified model. 

\section{Formulation of Problem}\label{problem}

An IDS is in essence an evaluator that can establish whether a set of network packets entering the system is malicious or not, by using what it has ``learned" from previous data. In order to successfully train an IDS for a bespoke system, a security professional needs to therefore collect large quantities of data. 
The problem with this is that to gather this data there are several options each with several drawbacks~\cite{mell2003overview}: 1) make use of known attack datasets to train the IDS 2) make use of existing IDS (e.g. Snort - Lightweight Intrusion
Detection for Networks~\cite{roesch1999snort}) or 3) make use of an exploit database and simulate attacks on your own system as a pen testing approach. 
This latter approach is by far the most precise~\cite{mell2003overview,bohme2010optimal} as it allows to search for bespoke attacks to the IoT network and construct a dataset which is unique and effective for the  specific system.

Whilst this approach produces the best suiting dataset it has some major drawbacks. 
Firstly, one must find and implement the attacks, which is a difficult process that might take a very long time\cite{mell2003overview}.
Secondly, one would need to cause major disruptions to one's own network by running the attacks which might obstruct work and productivity.
One of the many difficulties in detecting attacks on systems through the use of IDS is that one cannot (easily) predict potential attacker behaviour, or rather it is very difficult to classify an attack if its behavior differs from known attacks.  

We formulate the problem as the following: 
Firstly, is it possible to overcome some of these difficulties and train an IDS for a specific IoT system making use of a model?
The model would need to be able to produce similar results of the third approach, but would have the advantage that it could run parallel to the real system without causing downtime (Fig.1) .
Secondly, by making use of non-determinism and probabilistic behavior could the modelling approach recreate behavior that mimics that of an attacker probing and finding weaknesses in the system?
\begin{figure}[h]
  \label{fig:NIDS}
  \centering\includegraphics[width=\textwidth]{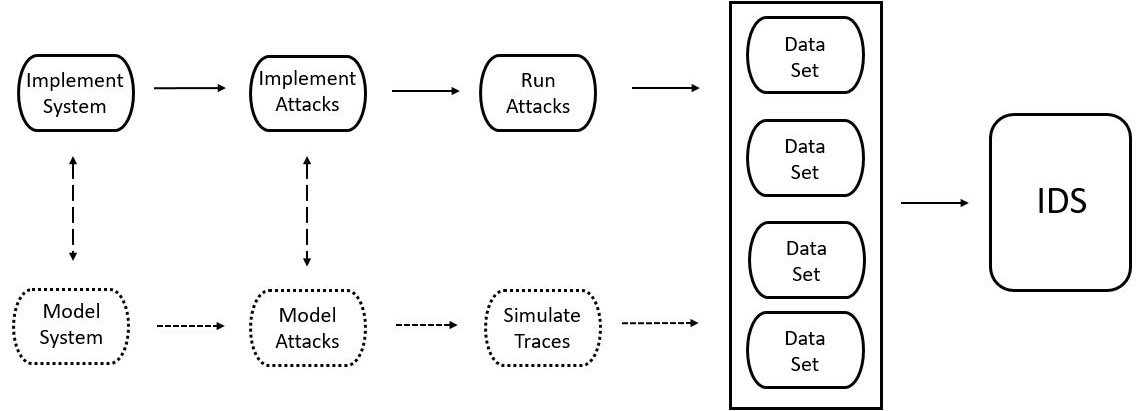}
   \caption{Running model along-side real system to generate further datasets}
\end{figure}

\section{IoT System Model} \label{IoT-system}

The intent of the modelled system is in essence to produce traces of behavior that correspond to the behaviour of real devices. A trace of a model under attack should be a subset of the full (finite) model trace. The traces are however limited by the drain of battery either by standard behavior or by attacker behavior, as devices out of battery stop performing actions.
This means that going from a set of traces one can reconstruct a data file of what has taken place in the system. The traces can be used by the IDS to observe patterns of behavior and set out rules to use against real-world attackers.

We model the system as a synchronization of three core components: a set of devices, a set of monitors (each assigned to a device) and an attacker.
The other aspect was measuring the impact of the actions on the system, specifically their effect on the devices battery and ability to operate successfully.
Whilst several process calculi are available to represent traces of processes we customize the trace semantics of standard process calculus, as there are several further features we need to capture in order to be able to produce descriptive datasets. 
Several process calculi achieve the notion of communication however this achieves the effect of two processes performing an action simultaneously, we wish to capture the effect of a device \emph{sending} an action and the other device \emph{receiving} it.
Looking at common process algebras such as CSP by Hoare \cite{hoare1978communicating}, the semantics of the traces are a set of action of the processes. These kinds of traces allow for a human reader to understand the way the system operates. However an IDS can draw very little information from these traces and they fail to capture the concept of {\em messages} through the network. 
Specifically we need to be able to capture the intercommunication between devices at each transition within the trace (as per a log in a real system).
This meant that output needed to hold further information rather than just the action taking place, the rules and the way the system was constructed was built around making a descriptive dataset.

The system as a whole is a tuple $\Phi = (D,M,T)$ where 
$D=\{D_1,..,D_n\}$  is a set of devices, $M=\{M_1,...,M_n\}$ is a set of monitors calculating properties of their corresponding device and $T=\{t_i,...,t_n\}$ is a set of times to calculate the changes over time of the system as a consequence of actions being triggered. We also introduce means to model an attacker as a malicious device.

\subsection{Device Model}\label{device-model}

Given a global set of actions $\gamma$, a device $D$ is a pair $(A,P)$, where $A \subseteq \gamma \times [0, 1]$ is the set of {\em active} actions, where $(a, p) \in A$ means the process chooses action $a$ with probability $p$, and such that $\sum \{p \mid (a, p) \in A\} = 1$; and $P \subseteq \gamma$ is the set of {\em passive} actions.

In order to recreate the full spectrum of potential system behaviours we model the set of actions of the device as the full capabilities of the real world device. 
This allows to capture the full set of abilities of its behavior and increases the accuracy of the benchmarking. 
This also eases the addition of further devices as they are simply modelled with the full send and receive action spectrum without the need to alter the rest of the system. 
The behavior of a device is in the form of a guarded communication, which in our model means that the communication is reliant on a set of conditions being true in order to be triggered. 
An action $a$ in the device can only be triggered to begin a communication if it doesn't violate the capabilities of the system, such as remaining battery and time per message. This allows for realistic device behaviour, mimicking the patterns and constrictions of a real system.

\subsection{Monitor Model}\label{monitor}

A monitor is the part of the system that enables its correct functioning as well as monitoring dangerous behavior. It calculates the shifts in battery of the various actions and synchronises with the devices to ensure correctness. 
A monitor $M$ controls value $\lambda$, where $\lambda$ is the remaining battery of the device.
Given a global set of battery drains $\Omega$ the $\lambda$ is measured as a quantity that is linearly drained by a $\omega_a$ where $\omega_a \in \Omega$ is a constant battery drain of an action, the monitor will update its $\lambda$ value to $\lambda'$ after each corresponding device action. The drain of each action is a fixed value calculated from the real world device, as such each action is associated to a single device only.

\begin{rul}
Given two devices $D_1 =(A_1, P_1)$ and $D_2 = (A_2, P_2)$, a communication initiated by device $D_1$ on an active action $a$, triggering corresponding receive action $\bar{a}$ in $D_2$, with an associated probability $p$ takes the form:
$$\frac{(a, p) \in A_1\ \ 
\bar{a}  \in P_2  
\ \ p > 0 
}
{(A_1, P_1) || (A_2, P_2) \xrightarrow{(a,p)}(A_1, P_1) || (A_2, P_2) } $$
\end{rul}

\begin{rul}
Given monitors $M_1$ and $M_2$ holding battery values $\lambda_1$ and $\lambda_2$, devices $D_1$ and $D_2$ are controlled by their respective monitors. 
The monitors calculate the drain in battery caused by action $a$ and $\bar{a}$ from constant drain values $\omega_a$ and $\omega_{\bar{a}}$ in the form:

\newcommand{\control}{\ensuremath{\vartriangleright}}
$$
\frac 
{
D_1 || D_2 \xrightarrow{(a,p)} D_1 || D_2 \quad
\lambda_1 > \omega_a \quad \lambda_2 > \omega_{\bar{a}}
}
{
 \lambda_1 \control D_1 || 
 \lambda_2 \control D_2 
\xrightarrow{(a, p)} 
{(\lambda_1 - \omega_a) \control D_1 || 
(\lambda_2 - \omega_{\bar{a}}) \control D_2}
}
$$
\end{rul}

These measurements can further aid the IDS in making informed decisions regarding the impact of the various actions in the system and were used to quantify the effectiveness of the attacker. 
Through this synthetic data the IDS will get a wide range of attacker behaviour that will lead to system failure, including potentially unknown attacker behaviour. 
Doing a similar approach without the model would require attacking one's own system and implementing an attack to collect data as per a penetration test (these approaches were compared in the experiments in section~\ref{experiments}).

\subsection{Traces of the system}

We differentiate each transition as a network packet running through the system, checked by the monitor of the device. Therefore they must be unique and fit all the possible behaviours of the device. As each action belongs to a single device it enables the corresponding devices to be uniquely identified.

\begin{rul}
\newcommand{\control}{\ensuremath{\vartriangleright}}
Given two devices controlled by their monitors in the form: $M_1\control D_1$ as $CD_1$ and $M_2 \control D_2$ as $CD_2$, and taking the total set of devices $X$, then the transition between CD1, CD2, taking system time {\em t} and being performed with probability {\em p} takes the form:
$$\frac{
CD_1||CD_2\xrightarrow{(a,p)}CD_1'||CD_2'\ \ 
}
{(t,CD_1||CD_2||X) \xrightarrow{(a,p,t)}(t + (t_a + t_{\bar{a}}),CD_1'||CD_2'||X)}$$
\end{rul}

In the computational view we compose a trace of the system inductively as a set of transitions in between states, where {\em prefix} is the prior transitions and the diagram describes a single transition in the form:\\

\begin{tikzpicture}[node distance=1cm]
\node (A) at (-1, 0)   {$\overbrace{\{\}}^\text{prefix}$};
\node (B) at (0.5, 0.1){$\overbrace{\bullet}^\text{state}$ };
\node (C) at (3.5, 0 ) {$\overbrace{[a,p,t']}^\text{transition}$ };
\node (D) at (7.5, 0.1){$\overbrace{\bullet}^\text{state'}$};
\node (I) at (0.5, -0.6) {$\downarrow$};
\node (K) at (3.5, -0.65){$\Updownarrow$};
\node (J) at (7.5, -0.6) {$\downarrow$};

\node (F) at (0.5, -1.5) {$
                            \begin{smallmatrix}
                              M_1 & D_1 & T_1\\ 
                              \vdots & \vdots & \vdots \\
                              M_n & D_n & T_n
                            \end{smallmatrix}
                       	 $};
\node (G) at (3.5, -1.5){$
                            \begin{smallmatrix}
                              \exists\ D_i\ \ni\ b_i\ >\ \omega_a\\
                              \ \wedge\ \ \exists\ D_j\ \ni\ b_j\  >\  \omega_{\bar{a}}\\
                              \wedge\ (a,p)\ \in \ D_i\ \wedge\ \bar{a}\ \in\ D_j 
                            \end{smallmatrix}
                          $};
\node (H) at (7.5, -1.5) {$
                            \begin{smallmatrix}
                               M_1' & T_1'\\ 
                               \vdots & \vdots  \\
                               M_n'& T_n'
                            \end{smallmatrix}
                            $
                            $
                            \begin{smallmatrix}
                              \forall\  k\ if\ k\in \{D_i,D_j\}  \\ 
                              \lambda_{k}'\mineq\; (\omega_a  + \omega_{\bar{a}})\ \\
                             \wedge \;t'\ \pluseq\ (t_a\; +\; t_{\bar{a}})\\
                              if\ k \not \in \{D_i,D_j\}\  M_k' = M_k
                            \end{smallmatrix}
                         $};
\end{tikzpicture}\\

The output of the system is a set of transitions following the semantics described. By generating the outputs of the system as the full behavior spectrum, the model can describe everything that can take place in the system. By updating the probabilities we can cater to the specifics of the underlying system behavior and make use of this to find unusual or potentially malicious behavior.
The rules can expand to include a wide array of behaviours and specifics to regulate devices actions and when they can be activated. These can include complex policies on whether actions can be activated at a specific time or whether some actions have higher priority allowing for very specific behaviour to be modelled.

\textbf{Running Example:} We show an example composed of: devices $D_x$,$D_y$ and $D_z$, corresponding monitors $M_x$,$M_y$ and $M_z$, and global time $t$. 
Each device has different actions that are synchronized with some other devices. 
The monitors have battery values for the devices and each device has a set $\Omega_i \in \Omega$ of action drains. 
Transitions follow the described rules to construct the traces. Note that they do not represent the full possible set of traces but rather two simulations of the system until devices are drained.  

\begin{table}
\label{running-example}
\centering
\caption{Example system model and its outputs}
\begin{tabular}{ll}
\hline
Devices : &  $\boldsymbol{D_x}=(A_x,P_x)$ where  \\
& $A_x = \{(read_{xy},0.3),(write_{xy},0.5),(read_{xz},0.2)\}$ and $P_x = \{ \overline{read_{zx}}\}$  \\
& $\boldsymbol{D_y}=(A_y,P_y)$ where \\  
& $A_y = \{(write_{yz},0.8),(read_{yz},0.2)\}$ and $P_y = \{\overline{read_{xy}}, \overline{write_{xy}},\overline{read_{zy}}\}$ \\
& $\boldsymbol{D_z}=(A_z,P_z)$ where \\ 
& $A_z = \{(read_{zx},0.1),(read_{zy},0.9)\}$ and $P_z = \{\overline{read_{xz}}, \overline{write_{yz}}, \overline{read_{yz}} \}$ \\
\hline

Monitors &  $\boldsymbol{M_x} \ni \lambda_x =5$ and $\boldsymbol{Drains_x}$ $\ni$ $\Omega_{A_x} = (1,3,1)$ $\wedge$ $\Omega_{P_x}=(1)$ \\
  \& Drains: &  $\boldsymbol{M_y} \ni \lambda_y =8$ and $\boldsymbol{Drains_y}$ $\ni$ $\Omega_{A_y} = (2,4)$ $\wedge$ $\Omega_{P_y}=(1,2,1)$ \\
&  $\boldsymbol{M_z}\ni  \lambda_z =2$ and $\boldsymbol{Drains_z}$ $\ni$ $\Omega_{A_z} = (1,1)$ $\wedge$ $\Omega_ {P_z}=(1,2,1)$ \\
\hline
Trace 1: & $[write_{yz},.8,30]$ 
$[write_{xy},.5,50]$ 
$[read_{xy},.3,65]$ 
\\
Trace 2: & 
$[read_{xz},.2,8]$ 
$[read_{xz},.2,16]$ 
$[read_{xy},.3,31]$ 
$[read_{xy},.3,46]$ 
$[read_{xy},.3,61]$\\
\hline                    
\end{tabular}
\end{table}

\subsection{Attacker Model}\label{attack-model}

An attacker synchronizes with a subset of actions of the device. When an attacker synchronizes on the device the monitor will synchronize on that action and calculate the respective drainage. 
The monitor keeps track of all these measurements for its respective device. 
Implementing the model in a tool like PRISM allows us to make use of Probabilistic Computation Tree Logic (PCTL)~\cite{baier2008principles} to calculate various conditions of pertinence to the system, to compute the optimal attack path, and to simulate traces of the model.

An attacker's intent is to behave in a manner that shortens the traces of the system by draining the value of battery in the monitor in the most efficient way possible.  
To model the attacker we made use of non-deterministic behavior in order to allow for anything to take place at any point. The advantage of non-determinism is that it allows for a system to arrive to an outcome using various routes. This can be manipulated to find optimal routes through the system and simulate varied behaviour. 
Unlike devices that are restricted by time and batteries of the devices they model, we allow for the attacker to have different levels of power to simulate various attacker strengths.  
An attacker, like the devices, has a set of unique actions $A_A$ a, however unlike other devices does not have a set of passive actions as it sits outside the connectivity of devices and cannot receive messages.
An Attacker may synchronise with any device in the system, and the set of actions $a_{A_i} \in A_A$ each correspond to different types of attacks in the real system. 
To expand further on the actions of the attacker, these should be very flexible and we make allowance for any action that can take place in the system (only restricted by the setup and protocols).

Each action label will correspond to an attack message from the real attacker and can be converted for the log file. 
For our specific example, each action in the attacker corresponds to the attacker in our experiment sending different packets/targeting different parts of the system as per the \textit{attacker experiment} in section~\ref{attack-example}.
Beyond actions it is important for us to be able to monitor the behaviour of attackers looking at how many actions an attacking device can perform at a time T (whether by assuming a real attacker device or by simulating different powers of attack). This is highlighted by measurements of the system we implemented that were then modelled in the monitor of each device. The other information to keep track of is: the choices the attacker makes to take down the devices, as these are important behavioural patterns for the IDS to use and can give us insight on potential vulnerabilities as well as unknown attacker behaviours.  

Unlike with the devices (whose intention is to cover the full spectrum of possible behaviours with the attacker), we are particularly interested in targeted behaviour. The attack actions therefore encompasses behaviours which are particularly damaging to the system (e.g. causes large drain of battery to the devices). As opposed to probabilistic behaviour we use non-determinism to find paths of behaviour that are particularly rewarding in terms of time taken to take down the system and in terms on lowering system usability (e.g. message throughput).
To model non-determinism we remove the probabilities from the attacker action. This differs from probabilistic behavior because the non-deterministic choice between process $A$ and $D_x$ is resolved at the moment the first action takes place. Conversely in the case of a probabilistic choice is done before the actions takes place~\cite{andova2002probabilistic}, so if there is a conflict in the system where both probabilistic actions and non-deterministic actions exist the probabilistic action is resolved first. 
By not associating a probability to an action we allow for the strategy of the attacker device to vary depending on what we are looking for in the system. 
Given a {\em policy} regulating the behaviour (corresponding to the available attack types) we allow for any action to take place at any point. This can be combined with a set of rules to find the trace of behaviour that allows to follow all the rules and yet still drain the battery as quickly as possible within these restrictions. 
Instead of a probability each action has an associated reward, and one can use this to find the path of most reward (or the best strategy to take down the system).

The non-determinism in combination with the reward structure $time$ is used to find the optimum \textit{attacker} strategy, or the most {\em rewarding} trace through the system. 
In PCTL it is written as $R\{``time"\}min=?[\ \textrm{F} \;  power=0 \ ]$ or the minimum time for the variable power (referring to battery levels) to reach 0. 
The value ``time" is a variable calculated by the time for a single message to be sent by the attacker and cumulated for each message sent before the power reaches zero calculated in microseconds and the power drain is calculated by the formulas in section~\ref{device-setup}. These reward structures allow for simulated attack strategies that an hypothetical attacker might make to take down the modelled IoT system.
Not all attacks rely on speed and intensity to take down the system, as highlighted by the running example in \ref{running-example} where the longer trace (Trace 2) is faster, so we model different rewards and observe different attacker behaviours. We can find generate traces of less detectable attack by associating an predictability score to an action and therefore keeping the behavior varied and realistic whilst still optimizing time. 
This can scale to several scenarios. We use these ``optimized" traces to create a large dataset that mimics different kinds of attackers.

\section{Experiment Methodology}\label{experiments}

To evaluate the effectiveness of the models we tested and compared the modelled system in~\ref{device-model} with the more standard approach described previously. Both the approaches output was used to train an IDS. The IDSs were then used to predict attack behavior. The verification was on the following basis: 1)  Accuracy on unknown attack detection; 2) Ability to mimic devices behavior and {\em smart} attackers. The setup of the experiment was the following:

\textbf{Experiment  - Device Setup : } \label{device-setup}
We set up a small IoT network in the lab and then modelled it to compare the results and to test out the effectiveness of our model in creating synthetic dataset.
For the sake of testing we kept the setup simple to display the tool as the thing that needs to scale and not the system. Once the simple model is created it is trivial to add more (similar) devices, whilst implementing a new system in the real world can be very time consuming.
We implemented a sensor network consisting of two devices. 
Each device had the following {\em actions}; they took sensor readings and then could \emph{send} it to the other device at any time; they could also \emph{request} the sensor data from the other device at any point. 
The devices used simple HTTP protocol for communication, and the behaviour was stored in Apache log format.
To accurately represent the devices and to create {\em smart} attackers, several measures needed to be obtained. Both devices were equipped with a Mh3500 battery.
We made a basic assumption that the devices are on constantly. We argue this is a correct assumption as due to our attack the device is constantly in log mode and therefore never in sleep mode.
Beyond this assumption we calculated time to send a message/log a message, baseline battery usage, percentage increase in battery usage under different DoS strains (taken this value and dividing it by messages processed for second) and battery drain per message. 

\textbf{Experiment - Attacker:}
\label{attack-example}
To validate the model we implemented a common DoS attack both in the real world and in the model. Our attack of choice was HULK, a DoS attacking tool which relies on several obfuscation techniques. In order to not be spotted whilst still outputting intense strain enough to take down systems very quickly~\cite{HULK2013}. The attack specifies it has the following properties:
1) obfuscation of source client - this is achieved by using a list of known user agents, and for every request that is constructed, the user agent is a random value out of the known list,
2) reference forgery - the referrer that points at the request is obfuscated and points into either the host itself or some major pre-listed websites,
3) stickiness - using some standard Http command to try and ask the server to maintain open connections by using Keep-Alive with variable time window and
4) unique transformation of URL - to eliminate caching and other optimization tools, they crafted custom parameter names and values and they are randomized and attached to each request.
The tool was able to take down a web server within minutes from just a single host. 
Seeing as IoT devices will have less capabilities than any web server we hypothesized that this would be a good attack to use as its properties make for a good dataset that is not straightforward to detect. These properties and obfuscations led to different combinations of message structure that we used in the non-deterministic attacker. 

To measure the time it takes per message we measure how many messages can be sent within a time period. This helps evaluate the accuracy in respect to the real world of our test attacker.
In order to measure voltage usage across the different IoT devices, we attached an extra component in between the battery supply and the device to take the readings required.
To measure battery drainage we utilized IoT battery lifespan estimator tool by Farnell~\cite{farnellelement14}.
This was used in combination with a variance we introduced on top of the calculator, to represent attack intensity and change to current. Through this we were able to estimate the different drains of the devices as an outcome of the actions they performed.
We created datasets utilising three approaches and compared each dataset in two different experiments.

The first dataset (RWD) was constructed from data from the real system. We implemented the system of devices and the real-world attack and monitored the behavior of the system. The data was logged across a period of twelve hours and used to train the first IDS. 
The second approach was a naive approach, we constructed a synthetic dataset (ND) without attacking the system but rather attempting random behavior. 
This gave a comparison of the model with a different synthetic dataset this will help evaluate the effectiveness of the IDS predictions as they effectively should be random guesses. 
And finally, we followed our proposed approach (MD) following section \ref{IoT-system}.

\subsection{Experiment 1}

As our dataset relies on stochastic events and actions, we created three datasets from the approach and evaluated each one to benchmark its effectiveness, a mean score was taken. Whilst our model is able to recreate very large datasets quickly we choose to keep the dataset size uniform across the initial experiment to get a fair comparison against the other two datasets. 
The comparison was based on accuracy of prediction against unknown attacks given IDSs trained with each of the datasets. The unknown dataset consisted of real world data of the systems behavior whilst being targeted by attacks that we had not modelled nor contained in the RWD. To measure accuracy we made use of the F score. The F score is a measure of a predictors accuracy, it is a measure of its precision over recall (a measure which takes in consideration both false positives and false negatives).

\subsection{Experiment 2}

The second experiment we ran was to test the effectiveness of the model in creating large quantities of behavior and the ability to readjust in case of network reconfiguration. We used deep learning classifiers catered to large datasets and created a much more efficient IDS purely through synthetic data. One of the core strengths of our approach is that once the model is setup the datasets are very easy to generate and we wanted to test whether this, in combination with our {\em smart} attackers, will lead to the ability to train better performing IDS.

\section{Experiment Setup}\label{setup}

To perform experiments described in section~\ref{experiments} we implement a Python framework that runs through the various steps required to test the IDSs: data generation, data processing, standardization and setting up of the IDS's classifiers.
This automatic framework prepares the datasets and trains the IDSs so that we may perform Experiment 1 and 2. It is implemented using the scikit-learn machine learning libraries.

\subsection{Data Generation}

Achieving a rich descriptive dataset was paramount in training an effective IDS. Through the outputted model traces  we were able to generate a dataset of different transitions through the modelled system. These traces were descriptive enough for a machine learning algorithms to construct rules about negative behavior through supervised learning. The traces of the model correspond to the real system behavior and each transition was labelled as either normal or abnormal behavior, therefore they can be used to make informed decisions about the system. For instance, if the model traces of the attacker continuously target a device, the IDS can interpret this as a weak point and set a rule to limit this behavior, as this could correspond to the behavior of a real world attacker.

\subsection{Data Processing \& Standardization}
To allow for  data to be interpreted by machine learning algorithms it needs to go through a process of standardization. 
This is often due to categorical non-numeric features or continuous features. The data provided by most if not all internet protocols is categorical (e.g. agent names and method calls). As such, in order to evaluate it we first needed to go through an initial phase of pre-processing. 
The intent of pre-processing is to render the data machine readable whilst preserving patterns. The process we adapted was the process of binarisation. 
Binarisation allocates a numeric value to each unique feature for example if dealing with HTTP codes GET would become 0001, POST 0010, DELETE 0100 and PUT 1000. 
This allows for the features to maintain their patterns and their predictive power and be used normally.
This initial step was applied to both the real world dataset and the naive synthetic dataset. This step was however not required for the model dataset as it already produced numeric features rather than categorical ones for efficiency.

\subsection{Classifiers}
The classifiers we implemented represented the IDSs. We choose to use two separate classifiers to get a better evaluation of the results. 
Each dataset was used to train two IDSs and then all the IDSs were tested against a new dataset of attack to establish their predictive power and the strength of the datasets.

The first classifier we implemented was Multi Layer Perceptron (MLP) Neural Network. 
An MLP consists of at least three layers of nodes. Except for the input nodes, each node is a neuron that uses a non-linear activation function~\cite{zhang2000neural}. 
MLP utilizes a supervised learning technique called back propagation for training.
Its multiple layers and non-linear activation distinguish MLP from a linear perceptron.
A linear perceptron is a function that can decide whether an input, represented by a vector of numbers, belongs to some specific class or not. Combining several together in an MLP and adjusting the functions and weights you build a statistically accurate classifier. The result is a non-linear perceptron that is able to classify non-linear classes. 

The second classifier used was a Decision Tree Classifier. A decision tree is a decision support tool that uses a tree-like graph or model of decisions and their possible consequences, including chance event outcomes, resource costs, and utility. It is one way to display an algorithm that only contains conditional control statements~\cite{safavian1991survey}.
Decision tree learning uses a decision tree (as a predictive model) to go from observations about an item (represented in the branches) to conclusions about the item's target value (represented in the leaves). 
The rules in the branches are automatically constructed from the training data which is labelled.
Using these rules it will be able to take in the test data and run it until it reaches an end node corresponding to a class (either DoS attack or normal behaviour).

\section{Results}\label{results}

Following the evaluation criteria in section~\ref{experiments} and recreating the model described in section~\ref{IoT-system}, we generated and tested three model datasets against our benchmarks of the naive dataset and the real world dataset. 
Beyond the accuracy of the results, we make an argument for feasibility and re usability of our approach.
The results were acquired by initially training two classifiers for each dataset, these were trained with 20,000 samples of which 10\% were attacks. The classifiers were then evaluated on an unknown and unlabelled real world dataset of 100,000 samples of which 20\% were attacks (of two different unknown types). The classifiers then attempted to label the new dataset to predict which ones were attacks.

\subsection{Experiment 1 - Results}
The neural network trained on the real world dataset proved to be very accurate with a 85.5\% prediction accuracy. On the other hand the model dataset trained predictor whilst still high, suffered from some degree of variance ($79.7 \pm 6.3 \%$).
What was of most interest however was the predictions outputted by the naive dataset of 0.9\%. This combined with the relatively inconsistent results of the synthetic dataset ($\pm 6.3\%$) make a case for over fitting. 
Over fitting is the scenario in which a model is trained so specifically to the training data that it is no longer classifying DoS attacks and normal behaviour of the system but rather focusing solely on the training data and learning on patters unique to the dataset not the system. 
This is quite common in Neural Networks as they perform best with very large quantities of data~\cite{zhang2000neural}, which for this part of the experiment we did not have. 

The results of the decision tree, contrasting to neural networks do not suffer from the same inadequacy of over fitting and do not necessarily need large amounts of data. 
This was mirrored by the results, as the model datasets all performed to very similar standards and the added randomness traces which might have disrupted the neural network made for a more ample rule set resulting in near perfect predicting power in the model dataset ($98.8 \pm 0.6$). 
The real world data which did not look at the possibility of random behaviour only achieved 77\% accuracy and the random dataset had a predictive power of near 50\% as expected.

\subsection{Experiment 2 - Results}
We observed that our approach of using non-determinism to recreate attack traces was particularly effective for the rule based classifier however led to disruption during the back-propagation process of the neural network, as non-standardized data can create uneven results.
This time using the much larger dataset of 100,000 transitions,the results were a lot more accurate (97.1\%) than previously, confirming our hypothesis.

As highlighted by this example our model has one key advantage over the traditional approach. Data generation is fast and efficient. 
If we wanted to improve the training of the IDS used on the real world dataset to a similar level of accuracy, it would take several days of data collection and consumption of resources (electricity, system downtime etc). 
We argue that whilst the initial effort of creating a model might be time consuming and perhaps not as intuitive for a potential system administrator, the phase of dataset generation makes up for this effort both for speed and predicting power of the IDS. 

\section{Conclusion \& Future work} \label{conclusion}

Our case study and proposed methodology has shown very promising results. We have shown that generating synthetic datasets of DoS attacks in IoT networks through this tool is both effective and efficient.  
We believe that the ability for this approach to scale easily to multiple devices and protocols in combination with its strong predictive power makes a very good argument for its usage across various IoT networks. 
Our argument for scalability of this approach is two fold, firstly it scales well in terms of costs as you can make assessment prior to implementing the system and secondly, we can bypass several of the downsides of verification (in terms of state space) as we focus on simulation. Perhaps the most useful feature of our proposed approach is that it allows for the construction of datasets to be very efficient even if a device is added or the system is reconfigured. As this is a prominent concern in dynamic IoT systems this advantage is quite significant.   

In this paper we included a case study of a single attack which worked very well. Our future work envisions the ability to model further attacks from a database to create an extensive set of attacks to create a much more predictive dataset.
We envision that the ability to relatively easily plug and play any IoT system in combination with implemented corpus of attacks, could turn into a tool that generates synthetic datasets of attacks to train bespoke IDSs for any IoT system.

\bibliographystyle{splncs03}
\bibliography{bibliography}

\begin{thebibliography}{10}
\providecommand{\url}[1]{\texttt{#1}}
\providecommand{\urlprefix}{URL }

\bibitem{farnellelement14}
Farnell element14, calculating battery life in iot applications (2017),
  \url{http://uk.farnell.com/calculating-battery-life-in-iot-applications}

\bibitem{HULK2013}
Hulk, web server dos tool - confessions of a dangerous mind (Feb 2013),
  \url{http://www.sectorix.com/2012/05/17/hulk-web-server-dos-tool/}

\bibitem{andova2002probabilistic}
Andova, S.: Probabilistic process algebra. Technische Universiteit Eindhoven
  (2002)

\bibitem{arnaboldi17dos}
Arnaboldi, L., Morisset, C.: Quantitative analysis of dos attacks and client
  puzzles in iot systems. In: Security and Trust Management {STM} 2017

\bibitem{baier2008principles}
Baier, C., Katoen, J.P., Larsen, K.G.: Principles of model checking. MIT press
  (2008)

\bibitem{bhuyan2014network}
Bhuyan, M.H., Bhattacharyya, D.K., Kalita, J.K.: Network anomaly detection:
  methods, systems and tools. IEEE communications surveys \& tutorials (2014)

\bibitem{bohme2010optimal}
B{\"o}hme, R., F{\'e}legyh{\'a}zi, M.: Optimal information security investment
  with penetration testing. In: International Conference on Decision and Game
  Theory for Security. pp. 21--37. Springer (2010)

\bibitem{buennemeyer2007battery}
Buennemeyer, T.K., Gora, M., Marchany, R.C., Tront, J.G.: Battery exhaustion
  attack detection with small handheld mobile computers. In: Portable
  Information Devices, 2007.

\bibitem{fruth2011formal}
Fruth, M.: Formal methods for the analysis of wireless network protocols.
  Oxford University (2011)

\bibitem{gubbi2013internet}
Gubbi, J., Buyya, R., Marusic, S., Palaniswami, M.: Internet of things (iot): A
  vision, architectural elements, and future directions. Future generation
  computer systems (2013)

\bibitem{guillen2015inefficiency}
Guillen, E., S{\'a}nchez, J., Paez, R.: Inefficiency of ids static anomaly
  detectors in real-world networks. Future Internet  7(2),  94--109 (2015)

\bibitem{hoare1978communicating}
Hoare, C.A.R.: Communicating sequential processes. Communications of the ACM
  21(8),  666--677 (1978)

\bibitem{kwiatkowska2002prism}
Kwiatkowska, M., Norman, G., Parker, D.: Prism: Probabilistic symbolic model
  checker. In: International Conference on Modelling Techniques and Tools for
  Computer Performance Evaluation (2002)

\bibitem{liang2016denial}
Liang, L., Zheng, K., Sheng, Q., Huang, X.: A denial of service attack method
  for an iot system. In: Information Technology in Medicine and Education
  (ITME), 2016 8th International Conference on. pp. 360--364. IEEE (2016)

\bibitem{mell2003overview}
Mell, P., Hu, V., Lippmann, R., Haines, J., Zissman, M.: An overview of issues
  in testing intrusion detection systems (2003)

\bibitem{Mirkovic:2004:IDS:1044905}
Mirkovic, J., Dietrich, S., Dittrich, D., Reiher, P.: Internet Denial of
  Service: Attack and Defense Mechanisms (Radia Perlman Computer Networking and
  Security). Prentice Hall PTR, Upper Saddle River, NJ, USA (2004)

\bibitem{mukkamala2002intrusion}
Mukkamala, S., Janoski, G., Sung, A.: Intrusion detection using neural networks
  and support vector machines. In: Neural Networks, 2002. IJCNN'02. Proceedings
  of the 2002 International Joint Conference on. vol.~2, pp. 1702--1707. IEEE
  (2002)

\bibitem{roesch1999snort}
Roesch, M., et~al.: Snort: Lightweight intrusion detection for networks. In:
  Lisa. vol.~99, pp. 229--238 (1999)

\bibitem{roman2013features}
Roman, R., Zhou, J., Lopez, J.: On the features and challenges of security and
  privacy in distributed internet of things. Computer Networks (2013)

\bibitem{safavian1991survey}
Safavian, S.R., Landgrebe, D.: A survey of decision tree classifier
  methodology. IEEE transactions on systems, man, and cybernetics  21(3),
  660--674 (1991)

\bibitem{exploitDB}
Security, O.: Exploitdb, offensive security's exploit database archive (2009),
  \url{https://www.exploit-db.com/}

\bibitem{suo2012security}
Suo, H., Wan, J., Zou, C., Liu, J.: Security in the internet of things: a
  review. In: Computer Science and Electronics Engineering (ICCSEE), 2012
  international conference on. vol.~3, pp. 648--651. IEEE (2012)

\bibitem{talpade2003mitigating}
Talpade, R., Madhani, S., Mouchtaris, P., Wong, L.: Mitigating denial of
  service attacks (Jan~29 2003), uS Patent App. 10/353,527

\bibitem{zhang2000neural}
Zhang, G.P.: Neural networks for classification: a survey. IEEE Transactions on
  Systems, Man, and Cybernetics, Part C (Applications and Reviews)  30(4),
  451--462 (2000)

\end{thebibliography}

\end{document}